\documentclass[%
 reprint,
 superscriptaddress,
 amsmath,amssymb,
]{revtex4-1}

\usepackage{graphicx}
\usepackage{dcolumn}
\usepackage{bm}
\usepackage{hyperref}
\usepackage{url}
\usepackage{comment}
\usepackage{xspace}
\usepackage{mathtools}

\newcommand{\eref}[1]{Eq.~(\ref{#1})}
\newcommand{\Eref}[1]{Equation~(\ref{#1})}
\newcommand{\fref}[1]{Fig.~\ref{#1}}
\newcommand{\Fref}[1]{Figure~\ref{#1}}

\newcommand{\appropto}{\mathrel{\vcenter{
  \offinterlineskip\halign{\hfil$##$\cr
    \propto\cr\noalign{\kern2pt}\sim\cr\noalign{\kern-2pt}}}}}

\newcommand\Ne{$n_\mathrm{e}$\xspace}
\newcommand\Te{$T_\mathrm{e}$\xspace}
\newcommand\Ti{$T_\mathrm{i}$\xspace}
\newcommand\NH{$n_\mathrm{H}$\xspace}
\newcommand\MFP{$l_\mathrm{MFP}$\xspace}

\newcommand{\mycomment}[1]{}

\begin{document}
\newcommand{\ManuscriptTitle}{
    A scaling law of the neutral opacity and Balmer-$\alpha$ wing shape in high-temperature plasmas 
}

\title{\ManuscriptTitle}

\author{Keisuke Fujii}
\email{fujiik@ornl.gov}
\affiliation{%
    Fusion Energy Division, Oak Ridge National Laboratory, Oak Ridge, TN 37831-6305, United States of America
}
\author{Masahiro Hasuo}
\affiliation{%
    Department of Mechanical Engineering and Science,
    Graduate School of Engineering, Kyoto University
    Kyoto 615-8540, Japan
}
\author{Motoshi Goto}
\affiliation{%
    National Institute for Fusion Science, Toki, Gifu, 5909-5292, Japan
}
\affiliation{%
    Department of Fusion Science, The Graduate University for Advanced Studies, SOKENDAI, Toki 509-5292, Japan
}
\author{Jeremy D. Lore}
\affiliation{%
    Fusion Energy Division, Oak Ridge National Laboratory, Oak Ridge, TN 37831-6305, United States of America
}
\date{\today}

\begin{abstract}
    Hydrogen atoms penetrating deep inside high-temperature magnetically confined plasmas by repetitive charge-exchange collisions result in a particle source, which affects the plasma performance significantly.
    In this \textit{Letter}, we present an approximate solution of the fluid equations for neutral transport and the analytical representation of the neutral opacity, in a simplified plasma geometry.
    This analysis predicts a power-law decay in the Balmer-$\alpha$ line wings which reflects the velocity distribution of the neutral atoms, with the power-law index analytically represented as well. 
    These scaling laws are validated by the comparison with a simple Monte-Carlo simulation and spectroscopic observations of Large Helical Device plasmas.
    Since the Balmer-$\alpha$ line wings are experimentally accessible, our formulation opens the door to directly observe the neutral opacity and thus the particle source distribution in the plasma.
\end{abstract}

\maketitle
\onecolumngrid

\textit{Introduction}

Hydrogen atoms penetrate deep inside high-temperature magnetically confined plasmas through repetition of charge exchange collisions, resulting in one of the particle sources of the plasma~\cite{Mordijck2020-br}.  %
Their dynamics significantly influences various aspects of the plasmas, e.g., particle, momentum, and heat transports, as well as the pedestal characteristics~\cite{Romanelli2015-yb,Mordijck2020-br}. 
Their penetration depth, sometimes called opacity, becomes more important in ITER~\cite{Kukushkin2011-ii} and future fusion plants since the neutral beam fueling will not be feasible in the core region because of the high density. 
Although it has been difficult to experimentally measure the neutrals deep in the plasma because the strong neutral emission at the far edge dominates over the emission from the inside, recently the authors have reported that the neutral atom emission from the core region can be clearly detected in the far wings of Balmer-$\alpha$ line~\cite{Goto2011-zu,Fujii2014a,Fujii2015-pp} for Large Helical Device (LHD). 
This method essentially utilizes the steep ion-temperature gradient around outside the pedestal, i.e., the spectrum is decomposed into multiple temperature components, where colder components represents the neutral at the outer region while hotter components represents those closer to the core region.
Now this measurement principle is used in other devices such as DIII-D and Wendelstein 7-X ~\cite{Haskey2022-cu,Winters2021-in}. 

As the governing processes of neutral transport are well known, numerical simulation methods have been established with various levels of fidelity, ranging from 2D and 3D full-kinetic Monte-Carlo simulations~\cite{Reiter1992-vq, Heifetz1982-hs} to more simplified 1D models~\cite{Tamor1981-tk,Hughes2006-ms,Fujii2013-ze}.
For example, a Monte-Carlo simulation code \texttt{EIRINE} solves the neutral dynamics from given plasma parameters, e.g., electron density \Ne and temperature \Te, ion temperature \Ti, the boundary conditions, such as the wall reflection, and the reaction rates of various elementary processes~\cite{Reiter1992-vq}.
Studies of the analytical dependence of the neutral transport on the plasma parameters have been also reported~\cite{Goldston1995-lm,Vold1990-dr,Mahdavi2003-rj}.
However, repetition of charge exchange collisions are often neglected in such analytical studies; the common assumption is that the neutral atoms experience just one charge exchange collision, which is not the case for high-temperature plasmas.

To understand the bidirectional neutral-plasma interaction and to aid in interpreting experimental observations, in this \textit{Letter}, we present a scaling law of the neutral opacity by taking the repetitive charge-exchange collisions into account.
This scaling law newly predicts a power-law decay in the Balmer-$\alpha$ line wings, the power-law index of which is analytically represented again by these parameters. 
We validate these scaling laws with a simple Monte-Carlo simulation as well as spectroscopic observations of LHD plasmas.
Since the Balmer-$\alpha$ line wings are experimentally accessible~\cite{Fujii2014a,Fujii2015-pp}, this opens the door to directly measure the neutral opacity by experiment.

\vspace{10mm}
\textit{Theory}


Previously, we have presented a fluid approximation of the neutrals, by taking repetitive charge-exchange collisions into account, i.e., the temperature change of neutrals during the transport is taken into account~\cite{Fujii2016-vh}.
The particle and energy balance of the neutrals can be written with the hydrogen density \NH and hydrogen temperature $T_\mathrm{H}$ as follows, respectively,
\begin{align}
    \label{eq:particle_balance}
    \nabla \cdot \left[\mu_\mathrm{H} \nabla (n_\mathrm{H}T_\mathrm{H})\right] 
    &= S_\mathrm{ion} n_\mathrm{e} n_\mathrm{H}\\
    \notag
    \frac{5}{2}\nabla \cdot \left[\mu_\mathrm{H} \nabla (n_\mathrm{H}T_\mathrm{H}^2)\right] 
    &= \frac{3}{2}\left(S_\mathrm{CX} + S_\mathrm{ion}\right) n_\mathrm{e} n_\mathrm{H} T_\mathrm{H} \\
    \label{eq:energy_balance}
    &-\frac{3}{2}S_\mathrm{CX} n_\mathrm{e} n_\mathrm{H} T_\mathrm{i}, 
\end{align}
$S_\mathrm{CX}$ is the charge-exchange rate coefficient normalized by the hydrogen-ion fraction in the plasma.
Thus, in hydrogen-dominant plasma, $S_\mathrm{CX} \propto Z_\mathrm{eff}^{-1}$, where $Z_\mathrm{eff}$ is the effective charge of the plasma particle.
Note that in this work, except for the following experiment-comparison section, we use the velocity squared unit for all the temperatures, so that the Boltzmann's constant and the atom mass are represented as unity. 
$\mu_\mathrm{H}$ is the mobility of hydrogen atoms, which can be written as 
\begin{align} 
    \mu_\mathrm{H} = \frac{1}{\left(S_\mathrm{CX} + S_\mathrm{ion}\right) n_\mathrm{e}}
\end{align} 
$S_\mathrm{ion}$ is the ionization rate coefficient. 
Since the temperature dependence of these rate coefficients are known, with given spatial distributions of \Ne, \Te, \Ti, and $Z_\mathrm{eff}$, and appropriate boundary conditions, Eqs \ref{eq:particle_balance} and \ref{eq:energy_balance} can be solved numerically to obtain the spatial distribution of \NH and $T_\mathrm{H}$~\cite{Fujii2016-vh}.

\begin{figure*}[tb]
    \includegraphics[width=18cm]{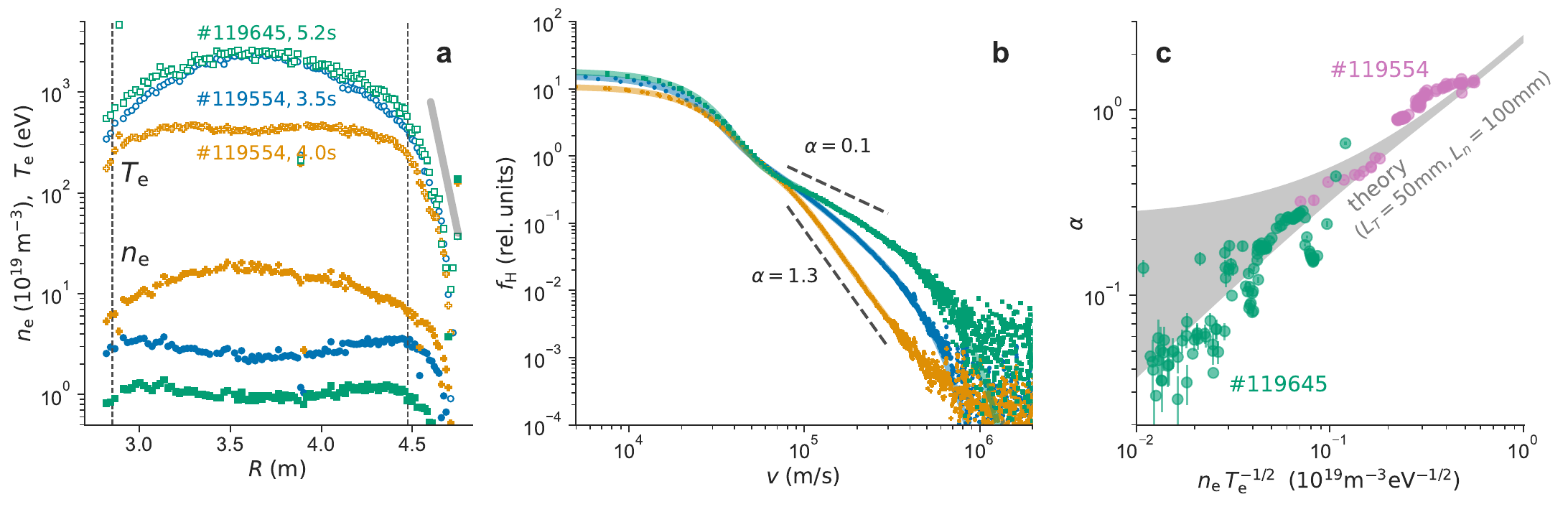}
    \caption{%
        (a)~Radial distributions of \Ne and \Te on the horizontally-elongated cross-section of the LHD.
        Three profiles at different experiment times are shown, which have different \Ne and \Te values.
        The thick gray line shows the exponential decay with $L_T=50$ mm.
        (b)~$f_\mathrm{H}$ observed for LHD plasmas for the three conditions. A power-law like wings can be seen, the power-law index of which are different depending on the plasma condition.
        (c)~The power-law index $\alpha$ of the spectrum for various plasma conditions as a function of $n_\mathrm{e}T_\mathrm{e}^{-1/2}$ (measured at the last closed flux surface).
        The theoretical predictions with $L_T=50$ mm and $L_n=100$ mm are shown by a gray shadow.
    }
    \label{fig:lhd_fH}
\end{figure*}

However here, instead of solving the equations numerically, we focus more on the parametric dependence of the solution by various simplifications. 
First, the plasma is assumed cylindrically symmetric with the radial coordinate $r$, and \Ne and \Ti are constant inside the pedestal top at $r_0$, and exponentially decaying profiles outside $r_0$,
\begin{align}
    \label{eq:density_profile}
    n_\mathrm{e}(r) &= \begin{cases}
        n_{\mathrm{e}0} & \quad (r < r_0) \\
        n_{\mathrm{e}0} \,e^{-(r - r_0) / L_n} & \quad (r \ge r_0) 
    \end{cases}\\
    \label{eq:temperature_profile}
    T_\mathrm{i}(r) &= \begin{cases}
        T_{\mathrm{i}0} & \quad (r < r_0) \\
        T_{\mathrm{i}0} \,e^{-(r - r_0) / L_T} & \quad (r \ge r_0) 
    \end{cases},
\end{align}
where the decay lengths are sufficiently small, $L_n \ll r_0$ and $L_T \ll r_0$.
As shown later in \fref{fig:lhd_fH}~(a), the scale lengths of the density and temperature decays are always much shorter (logarithmic gradient is larger) just outside the last closed flux surface than that of the inside.
Also, we neglect the temperature dependence of $S_\mathrm{ion}$ and $S_\mathrm{CX}$, and treat them as constants.
These coefficients indeed vary only $S_\mathrm{ion} = (1 \sim 3)\times 10^{-14}\mathrm{m^{-3}}$~\cite{Sawada1993-au} and $S_\mathrm{CX} = (1 \sim 10)\times 10^{-14}\mathrm{m^{-3}}$~\cite{international1999iaea} over the temperature range of $10^1 \sim 10^4$ eV.
$Z_\mathrm{eff}$ is assumed constant across the entire plasma.
These approximations make the system scale free and the ordering discussion easier, e.g., multiplying a constant $x$ to the spatial scale can be compensated by a $x^2$-times higher temperature, thus leading to the identical solution.

In this derivation, we particularly focus on the neutral transport at $r > r_0$, where the \Ne and \Ti has the strong gradients.
The two limiting cases will be considered depending on the (local) mean free path $l_\mathrm{MFP}$ of the charge-exchanged neutral
\begin{align}
    l_\mathrm{MFP} = \frac{1}{\left(S_\mathrm{CX} + S_\mathrm{ion}\right)} T_\mathrm{i}^{1/2}n_\mathrm{e}^{-1},
\end{align}
i.e., the long-\MFP limit ($l_\mathrm{MFP} \gtrsim L_n, L_T$) and the short-\MFP limit ($l_\mathrm{MFP} \ll L_n, L_T$).

At the short-\MFP limit, $T_\mathrm{H}$ becomes close to the equilibration with \Ti. 
Firstly, let us consider a special case for now, where $L_n = 2L_T$, which makes \MFP constant across the plasma.
Indeed, the smaller temperature-gradient scale length has been obtained in various devices~\cite{Sun2022-gc,Rudakov2005-qe}.
In this case, 
\begin{align}
    \label{eq:atom_density}
    n_\mathrm{H} = n_\mathrm{H0} \exp\left[\frac{r - r_0}{L_\mathrm{H}}\right]
\end{align} 
is the solution of \eref{eq:particle_balance}, which gives
\begin{align}
    \left(L_\mathrm{H}^{-1} + 4L_T^{-1}\right)\left(L_\mathrm{H}^{-1} + 2L_T^{-1}\right) = 
    S_\mathrm{ion} ( S_\mathrm{CX} +  S_\mathrm{ion}) n_\mathrm{e0}^2 T_\mathrm{i0}^{-1}.
\end{align}
Since we consider the short-\MFP limit, this may be simplified as 
\begin{align}
    \label{eq:atom_exponent}
    L_\mathrm{H}^{-1}\approx \frac{3}{4} L_T^{-1} + \sqrt{(S_\mathrm{CX} + S_\mathrm{ion})S_\mathrm{ion}} n_\mathrm{e} T_\mathrm{i}^{-1/2}.
\end{align}

This result can be extended to the other case $L_n \neq 2L_T$, by adopting the second order approximation, $n_\mathrm{H} = n_\mathrm{H0} \exp[L_\mathrm{H}^{-1} (r - r_0) + b (r - r_0)^2]$.
We obtain the identical result to \eref{eq:atom_exponent}.
Note that since the $L_\mathrm{H}$ is not constant for this case, its local value may be computed by the local values of \Ne and \Ti.
The detailed derivation will be described in a follow-on publication.

At the long-\MFP limit, while the system is affected by the boundary conditions more significantly and the fluid approximation is less valid, we find a rough estimate on the neutral transport.
As discussed in Refs.~\cite{Fujii2015-pp}, the hydrogen transport is nearly isobaric, i.e., the neutral pressure is almost constant across the plasma ($\nabla n_\mathrm{H} T_\mathrm{H} \approx 0$), while the atom temperature varies due to the charge-exchange collision.
This can be also found from the ordering consideration of \eref{eq:particle_balance}, which leads to 
$\nabla n_\mathrm{H} T_\mathrm{H} = \mathcal{O}(n_\mathrm{e}^{-2} T_\mathrm{i})$.
By integrating \eref{eq:energy_balance} with $\nabla n_\mathrm{H} T_\mathrm{H} = 0$ substituted, we obtain
\begin{align}
    \label{eq:energy_balance_largemfp}
    \mu_\mathrm{H} (n_\mathrm{H} T_\mathrm{H}) \frac{d}{dr} T_\mathrm{H}
    \approx -\frac{3}{5}S_\mathrm{CX} \frac{1}{r}\int_0^{r} n_\mathrm{e} n_\mathrm{H} \left(
         T_\mathrm{i} - \eta T_\mathrm{H}
    \right) rdr,
\end{align}
where $\eta = (S_\mathrm{CX} + S_\mathrm{ion}) / S_\mathrm{CX} \approx 1$.
We expect $T_\mathrm{H} \ll T_\mathrm{i}$ due to the large \MFP and short decay length-scales, the dominant heat flux of neutrals observed at the outside of the pedestal dominantly comes from the inside. Thus, right hand side (r.h.s.) of \eref{eq:energy_balance_largemfp} becomes constant at $r>r_0$, which leads to $T_\mathrm{H} \propto n_\mathrm{e}^{-1}$ there. 
Therefore, we again find $n_\mathrm{H} = n_\mathrm{H0} \exp[L_\mathrm{H}^{-1} (r - r_0)]$ with
\begin{align}
    \label{eq:atom_exponent_short}
    L_\mathrm{H} = L_n.
\end{align}
This behavior can be understood by the fact that the heating of the neutral by ions becomes exponentially more efficient as we go into the plasma, thus $n_\mathrm{H}$ decay has the same scale length to the plasma density.
We again note that as the \MFP is larger the boundary condition (e.g., the atom temperature outside the plasma) can affect to the solution more significantly.

\begin{figure}[tb]
    \includegraphics[width=12cm]{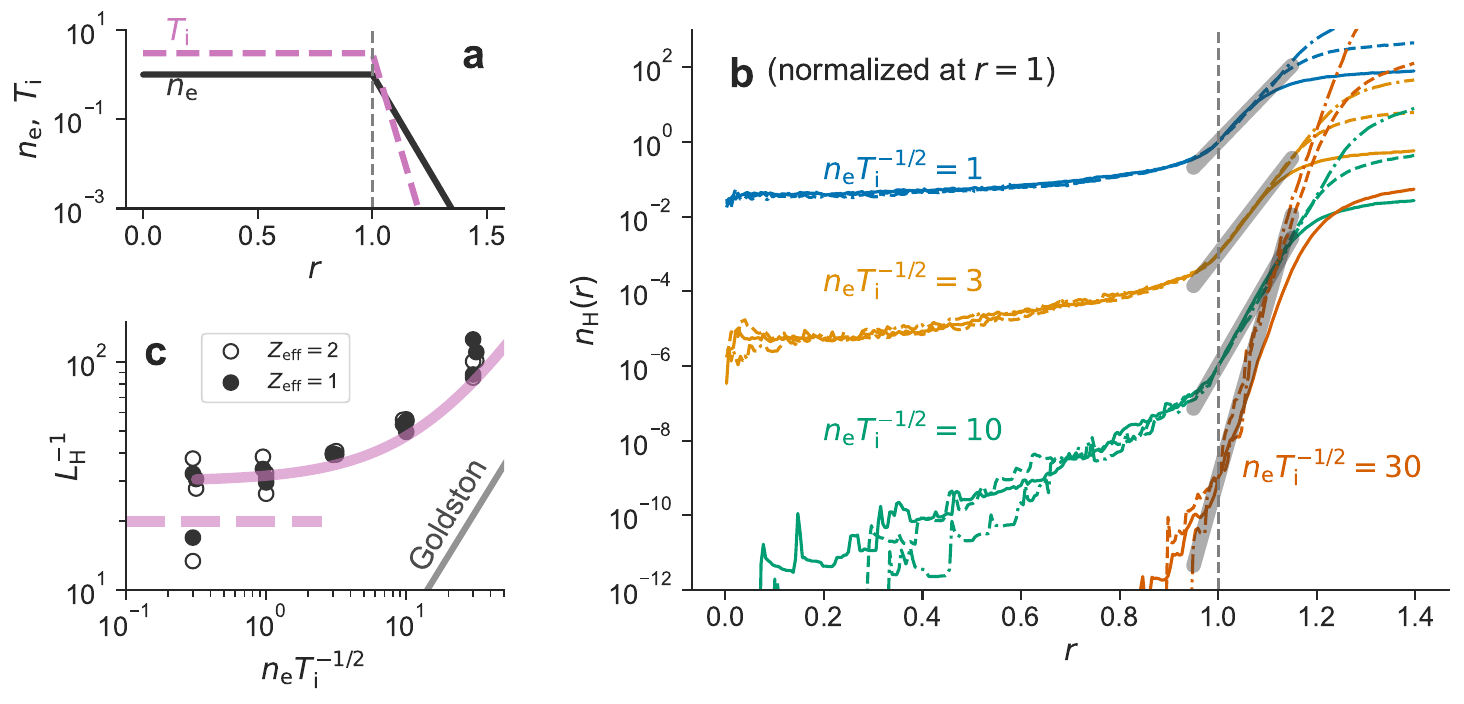}
    \caption{%
        A simple Monte-Carlo simulation for hydrogen transport.
        (a)~Radial distributions of \Ne and \Ti used for the simulation. The pedestal top is assumed at $r=1$.
        (b)~Simulated distributions of \NH for various values of \Ne and \Ti. The exponential decay is found at $r > r_0$.
        (c)~The scale length of the atom density as a function of \Ne and \Ti. The theoretical predictions \eref{eq:atom_exponent_short} and \eref{eq:atom_exponent} are shown by solid and dashed pink curves, respectively.
        The gray line shows the prediction by Goldston et al.~\cite{Goldston1995-lm}.
    }
    \label{fig:mc_nH}
\end{figure}

\begin{figure*}[tb]
    \includegraphics[width=17cm]{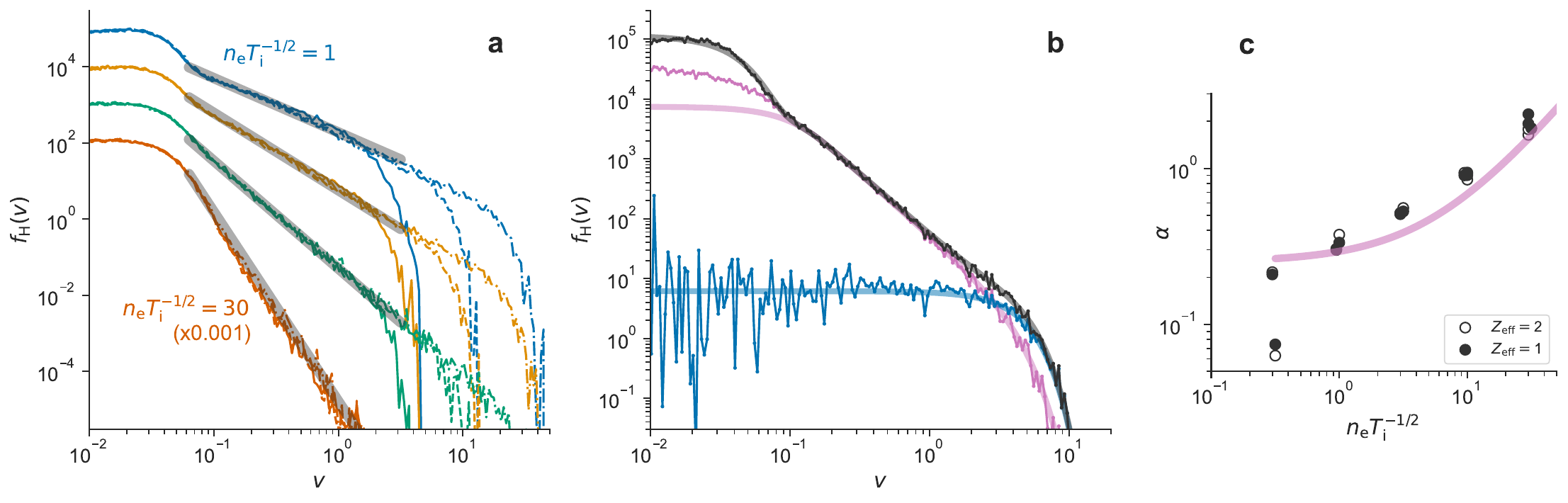}
    \caption{%
        Hydrogen velocity distribution obtained from the simple Monte-Carlo simulation.
        (a)~The velocity distribution for different plasma conditions. The power-law wing and its dependence on $n_\mathrm{e}T_\mathrm{i}^{-1/2}$ are clearly seen.
        (b)~Decomposition of the velocity distribution; pink markers shows the emission from neutrals that experience charge-exchange collision at $r > 1$ while blue markers are that at $r<1$.
        The solid gray curve shows the fit by \eref{eq:spectrum_total} to the simulated total distribution, while pink and blue curves show the components corresponding to $r > 1$ and $r < 1$, respectively, evaluated from the fit.
        (c)~The power-law index of the atom velocity distribution as a function of \Ne and \Ti. The theoretical predictions \eref{eq:power_index} is shown by a solid.
    }
    \label{fig:mc_fH}
\end{figure*}

Previously, another form of a diffusion equation has been proposed~\cite{Goldston1995-lm,Vold1990-dr}
\begin{align}
    \nabla \cdot \left(D_\mathrm{H}\nabla n_\mathrm{H}\right) = S_\mathrm{ion} n_\mathrm{e}n_\mathrm{H},
\end{align}
with 
\begin{align}
    D_\mathrm{H} = \frac{T_\mathrm{i}}{S_\mathrm{CX}n_\mathrm{e}}.
\end{align}
This model assumes the neutral temperature is identical to the ion temperature, which is similar to our short-\MFP limit, but also a spatially constant atom (ion) temperature across the plasma, which is in contrast to our approach (\eref{eq:energy_balance}).
This leads to a noticeable difference in the form; their temperature term is outside the first derivative (compare with \eref{eq:particle_balance}).
Instead, they solve the spatial profiles of $n_\mathrm{e}$ and $n_\mathrm{H}$ simultaneously, arriving at the same exponential form to \eref{eq:atom_density}, but with a different scale length~\cite{Goldston1995-lm,Mahdavi2003-rj}
\begin{align}
    \label{eq:atom_exponent_prev}
    L_\mathrm{H}^{-1}\approx \frac{1}{2} \sqrt{S_\mathrm{CX} S_\mathrm{ion}} n_\mathrm{e} T_\mathrm{i}^{-1/2}.
\end{align}
Comparing with our \eref{eq:atom_exponent}, at the short-\MFP limit, both the forms are proportional to $n_\mathrm{e} T_\mathrm{i}^{-1/2}$.
On the other hand, the above scale length does not have the offset term, and the proportional coefficients are different by a factor of $\approx 2$.


Let us consider the velocity distribution of neutral hydrogen observed as the Doppler broadening of the Balmer-$\alpha$ spectrum.
The hydrogen velocity distribution at a certain location $r$ can deviate significantly from a Maxwellian due to their large \MFP.
To compensate this effect, we take the same approximation proposed in Ref.~\cite{Fujii2015-pp,Fujii2016-vh}; 
an ensemble of atoms that are generated at $r$ through charge exchange collision well reflects the ion velocity distribution, which is close to a Maxwellian distribution, although the ensemble will diffuse out to a finite volume during their lifetime.
Such atoms will experience ionization collision or another charge exchange collision after traveling some distance, and during this travel the atoms may emit a Balmer-$\alpha$ photon with a certain probability. 
The emisssion probability $P_\alpha$ during one travel can be written by, with the photon-emission coefficient of Balmer-$\alpha$ line, $S_{\mathrm{H}\alpha}$,
\begin{align}
    P_\alpha \approx \frac{S_\mathrm{H\alpha}}{S_\mathrm{CX} + S_\mathrm{ion}},
\end{align}
which is approximately constant at $P_\alpha \approx 0.01$~\cite{Fujii2016-vh}.

Based on the constant $P_\alpha$ approximation, the volume integrated emission spectrum (with the wavelength shift from the rest wavelength $\Delta \lambda = \lambda - \lambda_0$ converted to the velocity along the sight line $v= -\Delta \lambda / c$) can be decomposed into the two components;
\begin{align}
    \label{eq:spectrum_total}
    f_\mathrm{H}(v) = f_\mathrm{mol}(v) + f_\mathrm{CX}(v),
\end{align}
where $f_\mathrm{mol}(v)$ corresponds to the emission from hydrogen atoms that do not yet experience a charge-exchange collision after recycled at the walls or puffed from the outside. This component should be low temperature $T_\mathrm{H}^\mathrm{mol}\approx 3$ eV.
$f_\mathrm{CX}(v)$ is the emission from the charge-exchanged neutrals, which can be derived by integrated over the observation volume $V$,
\begin{align}
    f_\mathrm{CX}(v) 
    &= \int_V P_\alpha S_\mathrm{CX} n_\mathrm{e} n_\mathrm{H} F(v, T_\mathrm{i}) dV\\
    \notag
    &= \cos\theta P_\alpha S_\mathrm{CX} n_\mathrm{e0} n_\mathrm{H0} 
     (2 T_\mathrm{i0})^\alpha v^{-2\alpha - 1} \\
    \notag
    &\times \frac{1}{\sqrt{\pi}} \left[
        \gamma\left(\alpha+\frac{1}{2}, \frac{v^2}{2T_\mathrm{i0}}\right)
        -\gamma\left(\alpha+\frac{1}{2}, \frac{v^2}{2T_\mathrm{i1}}\right)
    \right] \\
    \label{eq:spectrum}
    & + V P_\alpha S_\mathrm{CX} n_\mathrm{e0} \langle n_\mathrm{H} \rangle F(v, T_\mathrm{i0}),
\end{align}
with 
\begin{align}
    \label{eq:power_index}
    \alpha = (L_\mathrm{H}^{-1} - L_\mathrm{n}^{-1})L_\mathrm{T},
\end{align}
where $F(v, T)$ is the one-dimensional Maxwell distribution with temperature $T$, and $\gamma(s, x) = \int_0^x t^{s-1}e^{-t} dt$ is the lower incomplete gamma function.
$\theta$ is the angle between the sight line and the flux surface normal, assuming that $\theta$ is not too big so that  $\theta$ is constant over $r > r_0$. 
$\langle n_\mathrm{H} \rangle V$ is the total number of neutral atoms inside the pedestal and observation volume.
$T_\mathrm{i1}$ is the ion temperature at the place where the recycled hydrogen atom experiences the charge exchange-collision for the first time.

\Eref{eq:spectrum} tells that the hydrogen atom emission is represented by a truncated power-law distribution,
\begin{align}
    f_\mathrm{H}(v) \propto v^{-2\alpha -1}
    \;\;\;\;\;\;\;\; 
    (T_\mathrm{i1}^{1/2} \lesssim |v| \lesssim T_\mathrm{i0}^{1/2}).
\end{align}
This power law essentially originates from the combination of one exponentially increasing variable (temperature) and another exponentially decreasing variable (neutral density). 
This is a typical mathematical structure responsible for the emergence of power laws~\cite{Simkin2011-ab}.

\vspace{10mm}
\textit{Comparison with a Monte-Carlo simulation}


We test these analytical scaling laws with a simplified Monte-Carlo simulation.
In this simulation, a cylindrically symmetric plasma with the profiles Eq.~\ref{eq:density_profile}--\ref{eq:temperature_profile} (\fref{fig:mc_nH}~(a)), and neutral atom transport are tracked from the recycled outside the plasma. 
The atom temperature recycled from the wall is assumed to be 0.001, while $r_0 = 1$, $S_\mathrm{ion} = 1$, $S_\mathrm{CX} = 2 Z_\mathrm{eff}^{-1}$, and $S_\mathrm{H\alpha} = 1$ are used.
Also, we assume $L_\mathrm{n} = 0.05$ and $L_\mathrm{T} = 0.025$, thus $L_\mathrm{n} = 2L_\mathrm{T}$. 
Although all the governing processes are scale free, these numbers are chosen from the realistic scales of typical fusion experiments, i.e., the lengths are in the unit of 1 m, the temperatures are in 1 keV, the densities are in $\mathrm{10^{19}\;m^{-3}}$, and the rate coefficients are in $10^{-14}\;\mathrm{m^3}$.

\Fref{fig:mc_nH}~(b) shows the simulated spatial distributions of the neutral for several values of \Ne and \Ti.
Exponential decay in \NH is found in the distribution at $r \approx r_0$, and the scale length $L_\mathrm{H}$ changes depending on $n_\mathrm{e} T_\mathrm{i}^{-1/2}$.

\Fref{fig:mc_nH}~(c) shows the density and temperature dependence of the neutral scale length $L_\mathrm{H}$. 
The scale length depends on $n_\mathrm{e}T_\mathrm{i}^{-1/2}$, as expected from theories, particularly at the short-\MFP region (large $n_\mathrm{e}T_\mathrm{i}^{-1/2}$ side).
At the long-\MFP region, since the boundary effect becomes more important and thus the scatter becomes bigger.
The solid and dashed pink curves correspond to our theoretical predictions, \eref{eq:atom_exponent} and \eref{eq:atom_exponent_short}.
Also, by a gray solid line we show the theory by Goldston et al.~\cite{Goldston1995-lm}.
Our prediction shows a good consistency with the simulation, particularly at the short-\MFP region.
Note that as the theory by Goldston et. al does not specify what ion temperature to be used (currently we use the pedestal top temperature), the direct comparison may be difficult, e.g., with the lower ion temperature this theory will give a closer result to the simulation.

\Fref{fig:mc_fH} shows the simulated results of the velocity distribution of the neutral hydrogen integrated over the plasma obtained from Balmer-$\alpha$ spectrum.
Note that we do not assume the constant $P_\alpha$ in the Monte-Carlo simulation, but uses the constant $S_{\mathrm{H}\alpha}$.
The power-law decay is seen in the spectral wing, the power-law index of which depends on $n_\mathrm{e} T_\mathrm{i}^{-1/2}$ but independent from the ratio between them, which are again consistent with the discussion above.

\Fref{fig:mc_fH}~(b) shows components in the integrated velocity distribution; the emission from atoms which the last charge exchange collision they experience are $r > r_0$ (pink marker) and that from $r < r_0$ (blue markers).
The gray solid curve is a fit to the total spectrum (black markers) by \eref{eq:spectrum_total} assuming that $n_\mathrm{mol}$, $n_\mathrm{H0}$, $\langle n_\mathrm{H} \rangle$, $T_\mathrm{i0}$, $T_\mathrm{i1}$, and $\alpha$ are adjustable.
This well represents the total spectrum.
The first and second terms of \eref{eq:spectrum} are shown by solid pink and blue curves, respectively.
Each component is correctly captured by the fit to the total spectrum, particularly at the high velocity side.
It suggests the validity of our derivation.

\Fref{fig:mc_fH}~(c) shows the parameter dependence of $\alpha$, which is evaluated from the fit process to the total spectrum.
This is consistent with our theoretical prediction (pink curve), \eref{eq:power_index} with \eref{eq:atom_exponent_short} substituted. 

\vspace{10mm}
\textit{Experimental Comparison}


The volume-integrated velocity distribution $f_\mathrm{H}(v)$ is experimentally accessible.
\Fref{fig:lhd_fH}~(b) shows $f_\mathrm{H}$ in Large-Helical-Device (LHD) plasma observed from the Balmer-$\alpha$ spectrum.
The experimental apparatus is the same to that described in Refs.~\cite{Fujii2015-pp,Fujii2016-vh}, where the emission from an entire poloidal cross section of the LHD was collected and measured by a high-dynamic-range spectrometer~\cite{Fujii2014a}.
The three spectra are shown in the figure, which are taken at three different densities and temperatures. 
The radial profiles of the electron density and temperature at the three timings are shown in \fref{fig:lhd_fH}~(a).
The experimentally observed $f_\mathrm{H}$ shows the power-law dependence in its wings (as indicated by dashed lines in \fref{fig:lhd_fH}~(b)), and its power-law index changes according to the plasma parameters.

The spectra are fitted as shown by the bold curves in \fref{fig:lhd_fH}~(b).
The spectral shapes are well reproduced by the theoretical model, although drastic approximations are adopted in the theory.

\Fref{fig:lhd_fH}~(c) shows the power-law exponent $\alpha$ obtained from the fit for series of plasma experiments, as a function of $n_\mathrm{e}T_\mathrm{e}^{-1/2}$ observed at the last closed flux surface ($R\approx 2.85, 4.48\;\mathrm{m}$).
Note that as the ion temperature data is not always available particularly at the edge, we assume $T_\mathrm{e} \approx T_\mathrm{i}$.
This assumption might be valid for the edge region, due to higher collisionality there.
The positive dependence of $\alpha$ on $n_\mathrm{e}T_\mathrm{e}^{-1/2}$ is clearly seen.
The theoretical prediction of $\alpha$ assuming $L_n = 100\;\mathrm{mm}$ and $L_T = 50\;\mathrm{mm}$ (see the gray line in \fref{fig:lhd_fH}~(a)) are shown by the gray shadow, the uncertainty area of which comes from the two limiting cases \eref{eq:atom_exponent} and \eref{eq:atom_exponent_short}.
$S_\mathrm{ion} = 3\times 10^{-14} \;\mathrm{m^3/s}$ and $S_\mathrm{CX} = 3\times 10^{-14} \;\mathrm{m^3/s}$ are used.
This prediction well reproduces the experimental results.

We note that $L_n$ and $L_T$ indicate the scale lengths of the electron density and temperature decays, respectively, at the \textit{penetration position} of the neutral atoms.
Since these scale lengths has a poloidal dependence (in LHD the length scale is larger at the X-point, where the Thomson scattering was made in \fref{fig:lhd_fH}~(a)), the effective values of $L_n$ and $L_T$ may be affected by the poloidal distribution of neutrals.
The discrepancy of $\alpha$ from the theory that is seen in \fref{fig:lhd_fH}~(c) can reflect such changes in the neutral transport, as well as the change in the electron density and temperature distributions.
Currently, the volume integrated spectrum is only considered. 
However, in principle the spatial distribution of the Balmer-$\alpha$ spectrum can be measured by multiple sight lines, up to the resolution of $\approx l_\mathrm{MFP}$.
Such a multi-sight-line measurement of $\alpha$ may provide the poloidal dependence of $L_T$, which has been difficult by conventional methods.

\vspace{10mm}
\textit{Conclusion}

From the fluid approximation of the neutral transport and simplified plasma geometry, we newly derived the scaling law of the neutral opacity.
This theory predicts the power-law decay in the Balmer-$\alpha$ wings.
We confirmed our scaling law by comparing with a Monte-Carlo simulation and the Balmer-$\alpha$ spectrum observed for LHD plasmas.
Since the Balmer-$\alpha$ spectrum is experimentally accessible, our new derivation may be useful to study the neutral transport and the particle source in the plasma.

\vspace{10mm}
\textit{Data Availability}

All the experimental data used in this work can be found in the LHD experiment data repository \url{https://www-lhd.nifs.ac.jp/pub/Repository_en.html}.

\begin{acknowledgments}
    This work was partly supported by the U.S. D.O.E contract DE-AC05-00OR22725. 
    K.F. would like to specifically acknowledge Oak Ridge National Laboratory's Laboratory Directed Research and Development program Project No. 11367.
    K.F. thanks Dr. Shaun Haskey at Princeton Plasma Physics Laboratory, and Dr. Bart Lomanowski and Dr. Jae-Sun Park at Oak Ridge National Laboratory for fruitful comments and feedbacks to this work.
\end{acknowledgments}

\renewcommand{\thefigure}{A\arabic{figure}}
\setcounter{figure}{0}
\renewcommand{\thetable}{A\arabic{table}}
\setcounter{table}{0}
\renewcommand{\theequation}{A\arabic{equation}}
\setcounter{equation}{0}


\bibliography{refs}

    


\begin{widetext}
Notice:  This manuscript has been authored by UT-Battelle, LLC, under contract DE-AC05-00OR22725 with the US Department of Energy (DOE). The US government retains and the publisher, by accepting the article for publication, acknowledges that the US government retains a nonexclusive, paid-up, irrevocable, worldwide license to publish or reproduce the published form of this manuscript, or allow others to do so, for US government purposes. DOE will provide public access to these results of federally sponsored research in accordance with the DOE Public Access Plan (\url{http://energy.gov/downloads/doe-public-access-plan}).
\end{widetext}

\end{document}